\documentclass[review]{elsarticle}

\usepackage{lineno,hyperref}
\modulolinenumbers[5]

\usepackage{graphicx,epsfig}
\usepackage{times,bbm}
\usepackage{graphics,dcolumn,bm,float}
\usepackage{amssymb,amsmath,rotate,color}
\usepackage[title,titletoc,toc]{appendix}
\usepackage{mathtools}
\usepackage{booktabs}
\usepackage{tcolorbox}

\usepackage{wrapfig}
\usepackage{lipsum}
\usepackage{mwe}
\usepackage{braket}
\usepackage{physics}
\usepackage[caption=false]{subfig}
\usepackage[export]{adjustbox}
\usepackage{tikz}
\usetikzlibrary{through,calc}
\usetikzlibrary{positioning}

\begin{document}
\unitlength 1.5 cm
\newcommand{\be}{\begin{equation}}
\newcommand{\ee}{\end{equation}}
\newcommand{\bea}{\begin{eqnarray}}
\newcommand{\eea}{\end{eqnarray}}
\newcommand{\nn}{\nonumber}
\newcommand{\hlt}[1]{{#1}}
\newcommand{\up}{\uparrow}
\newcommand{\down}{\downarrow}
	
\newcommand{\bl}[1]{\textcolor{blue}{#1}}
\newcommand{\re}[1]{\textcolor{red}{#1}}
\newcommand{\gr}[1]{\textcolor{green}{#1}}

\newcommand{\bss}{{\boldsymbol{s}}}
\newcommand{\bsp}{{\boldsymbol{p}}}
\newcommand{\bsq}{{\boldsymbol{q}}}
\newcommand{\bsk}{{\boldsymbol{k}}}
\newcommand{\bsX}{{\boldsymbol{X}}}
\newcommand{\bseta}{{\boldsymbol{\eta}}}
\newcommand{\bsSigma}{{\boldsymbol{\Sigma}}}
\newcommand{\pr}{\partial}
\newcommand{\bs}{\boldsymbol}
\newcommand{\bearr}{\begin{eqnarray}}
\newcommand{\eearr}{\end{eqnarray}}
\begin{frontmatter}

\title{Generalization of Lieb-Wu wave function inspired by one-dimensional ionic Hubbard model}

\author[mymainaddress]{Abolfath Hosseinzadeh}

\author[mymainaddress]{S.A. Jafari\corref{mycorrespondingauthor}}
\cortext[mycorrespondingauthor]{Corresponding author}
\ead{jafari@physics.sharif.edu}

\address[mymainaddress]{Department of Physics$,$ Sharif University of Technology$,$ Tehran 11155-9161$,$ Iran}

\begin{abstract}
With the ionic Hubbard model (IHM) in mind, we construct a non-trivial generalization of the Bethe ansatz (BA) wave function which naturally 
generalizes the Lieb-Wu wave function with an ionic parameter $\Delta$, and reduces to Lieb-Wu solution in the limit $\Delta\to 0$. 
The resulting two-particle scattering matrix satisfies the Yang-Baxter equation.
To the extent that the unit cells with more than two electrons (Choy-Haldane issue) are avoided on average, 
our wave function represents an effective soluiton for the one-dimensional IHM. 
The Choy-Haldane issue limits the validity of our solution to low-filling and large $U\gtrsim 4$.
This regime is attainable in cold atom realizations of the IHM. For this regime, we numerically solve the 
generalized Bethe equations and compute the ground state energy in the thermodynamic limit. 
\end{abstract}


\end{frontmatter}


\section{INTRODUCTION}
Bethe Ansatz (BA) is a powerful method to construct exact wave functions for the ground and excited states 
of vast classes of one-dimensional Hamiltonians~\cite{article,Levkovich_Maslyuk_2016,Yupeng_Wang_2015,essler_frahm_göhmann_klümper_korepin_2005,doi:10.1142/2148,PhysRevLett.74.4507,PhysRevLett.96.216802,PhysRevB.96.115424,PhysRevLett.43.1698,PhysRevLett.79.1901,PhysRevLett.52.364,gaudin_2014,RevModPhys.55.331,Tsvelick_1983}. 
It provides solutions of Heisenberg model~\cite{Bethe1931}, Lieb-Liniger Model~\cite{PhysRev.130.1605}, Hubbard model~\cite{PhysRevLett.21.192.2, LIEB20031}, 
XXZ~\cite{PhysRev.112.309,PhysRev.150.321}, XYZ~\cite{PhysRevLett.26.834,BAXTER1972323}, Temperley-Lieb spin chain~\cite{NEPOMECHIE2016910}, Thirring model~\cite{PhysRevLett.42.135,PhysRevD.19.3666} and Chiral-Invariant Gross-Neveu Hamiltonian ~\cite{PhysRevLett.43.1698}. 

\begin{figure}[t]
\centering
\includegraphics[scale=0.4]{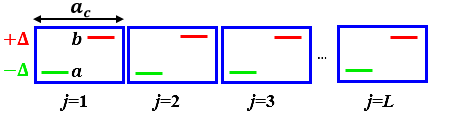}
\caption{\label{figure1}One-dimensional ionic Hubbard model energy schematic by considering two orbitals in each unit cell,
each belonging to one sublattice. \hlt{We set the unit cell length by $a_c=1$.}} 
\end{figure}

Among all these, 
the celebrated Hubbard model addresses the competition between kinetic energy and the on-site
Coulomb repulsion between electrons of opposite spins~\cite{doi:10.1142/2945,hubbard1981model}. An interesting deformation
of this model is the so called ionic Hubbard model~\cite{Egami1307}. The IHM is the Hubbard model 
plus an additional alternating (ionic) scalar potential of strength $\Delta$ (see Fig.~\ref{figure1}).
The essential phyiscs of this model is the competition between the inoic potential $\Delta$ and the
Hubbard term $U$. This model in arbitrary dimension and at half-filling has been the subject of extensive studies
with various techniques~\cite{Fabrizio1999,Martin2001QMC1D,kampf2003bondOrder,gros2003minimal,manmana2004quantum,Aligia2004Charge,Aligia2005Dimerized,Jabben2005RG,
maeshima2005optical,Mancini2005Composite,Solyom2006Entropy,Garg2006,Bouadim2007,craco2008dmft,hafez2009flow,McKenzie2009Triangular,Kampf2009Order,
Baeriswyl2009Critical,HafezJafari2010EPJB,HafezAdibi2010,hoang2010metal,HafezAbolhassani2011,Go2011Phase,
kim2012QMC,ebrahimkhas2012,Garg2015,Hafez2014,Esslinger2015Ultracold,Hafez2016,adibi2016,Kollath2017spectroscopy,
Farajollahpour2017}:
In infinite dimension the method of dynamical mean field theory (DMFT) has been applied to investigate the phase transitions of the IHM 
at half filling~\cite{Garg2006,Garg2015,ebrahimkhas2012,craco2008dmft,Kampf2009Order,Jabben2005RG,Go2011Phase}. 
For large $U$ ($\Delta$) the Mott (band insulating) phase is stabilized.
While with cellular DMFT one obtains a direct transition between Mott and band insulating phases~\cite{Go2011Phase}, 
within a single-site DMFT one obtains an intermediate conducting phase, i.e. when $U$ and $\Delta$ become comparable, the parent conducting states 
is restored~\cite{Garg2006,Kampf2009Order,ebrahimkhas2012}. 
This agrees with coherent potential treatment of Hoang~\cite{hoang2010metal}. 
A similar picture can be obtained from continuous unitary transformations~\cite{hafez2009flow}. This method has been extensively 
applied by Hafez and co-workers to study the excitaiton sepctrum of this model~\cite{HafezJafari2010EPJB,HafezAdibi2010,HafezJafari2010EPJB,
HafezAbolhassani2011,Hafez2014,Hafez2016}.

In two space dimensions the picture remains unclear. Determinantal quantum Monte Carlo study gives a metallic
phase~\cite{Bouadim2007}. 
Building on an orthogonal metallic state~\cite{SenthilOM}, interesting possibility 
of semiconductor of spinons has been proposed~\cite{Farajollahpour2017}. 
On triangular lattice, crossovers between Mott and charge transfer and covalent insulators and magnetically ordered
states is reported~\cite{McKenzie2009Triangular}. 
Aligia used charge and bond operator formalism to study the phase diagram of IHM in various dimensions~\cite{Aligia2004Charge}.
Mancini has employed his composit operator method to study IHM~\cite{Mancini2005Composite}. 

In one dimension, Fabrizio and co-workers used the bosonization to address the competition between $U$ and $\Delta$
at half-filling. They find an spontaneously dimerized phase~\cite{Aligia2005Dimerized} between the Mott and band insulating phase~\cite{Fabrizio1999}. 
This agrees with numerical calculations of Otsuka and Nakamura based on level-crossing and renormalization group
method based on exact diagonalization~\cite{Nakamura2005}.
IHM in 1D was subsequently studied with density matrix renormalization group (DMRG)~\cite{kampf2003bondOrder,manmana2004quantum,Baeriswyl2009Critical}
and Quantum Monte Carlo~\cite{Martin2001QMC1D,kim2012QMC} which support the bond ordered state. 
This model has also been studied using an interesting two-site entropy as a diagnosis tool of 
phase transitions~\cite{Solyom2006Entropy}. The opitcal response of an extension of IHM has also 
been studied by Maeshima and Yonemitsu~\cite{maeshima2005optical}. 

Given the recent cold atom realization of IHM~\cite{Esslinger2015Ultracold} and the invention of superlattice
modulation spectroscopy to probe the bond order~\cite{Kollath2017spectroscopy}, and a very appealing 
recent proposal to realized 1D IHM in twisted bilayer GeSe~\cite{Dante2019},
an exact solution of this model in 1D would be desirable. 
The first question one faces in this direction is the quantum integrability of the model.
In our previous work~\cite{hosseinzadeh2019integrability}, using energy level spacing statistics as well as the eigenvalues of the one-particle density matrix,
we have established the many-body localization for IHM. Therefore, there should exist an exact solution.
However the Choy-Haldane issue~\cite{Choy_1982} prevents the Bethe ansatz from solving the IHM:
Beyond two electrons per unit cell (containing two sites or "orbitals" with energies $\pm\Delta$) in Fig.~\ref{figure1}, 
the BA wave function will not be eigen-state of the IHM~\cite{Choy_1982,LEE1990398,Frahm_1993}. 
The essence of Choy-Haldane issue is that, there are configurations in the Hilbert space of the
actual IHM that are not accessible by Bethe ansatz wave functions. 
In the limit of large Hubbard $U$ which prevents the double occupancy of every orbital, 
the Choy-Haldane issue is "effectively" avoided at low-filling. Furthermore, in the $\Delta=0$ limit, there is no Choy-Haldane issue.
Therefore it is hoped that the Bethe ansatz wave function is a reasonable approximation of the exact eigen functions of the IHM in the 
following two regimes: (i) small $\Delta$ and arbitrary filling and (ii) large $U$ and low filling and the 
Choy-Haldane issue will causes parametrically small effects in the above regimes.

\section{MODEL HAMILTONIAN}
Hamiltonian describing the one-dimensional IHM, is an innocent-looking extension of the Hubbard model in which the lattice sites 
are assigned alternating electrostatic potentials $\pm\Delta$ as shown in Fig.~\ref{figure1}.
The Hamiltonian of IHM is given by
\begingroup\makeatletter\def\f@size{10.5}\check@mathfonts
\begin{eqnarray}
H_{\rm IHM}=H_{\rm t}+H_{\rm U}+H_{\Delta},\label{IHM.eqn}
\end{eqnarray}
\endgroup
where $H_{\rm t}+H_{\rm U}$ and $H_{\Delta}$ are the standard Hubbard Hamiltonian and the ionic part respectively which are described by,
\begingroup\makeatletter\def\f@size{10.5}\check@mathfonts
\begin{subequations}
\begin{eqnarray}
H_{\rm t}=&&-t\sum_{j=1,\sigma}^{L}\left( c_{j,\sigma}^{a^{\dagger}}c_{j,\sigma}^{b}+c_{j+1,\sigma}^{a^{\dagger}}c_{j,\sigma}^{b}+c_{j,\sigma}^{b^{\dagger}}c_{j,\sigma}^{a}+c_{j,\sigma}^{b^{\dagger}}c_{j+1,\sigma}^{a}\right) \\
H_{\rm U}=&&U\sum_{j=1}^{L}\left(c_{j,\uparrow}^{b^{\dagger}}c_{j,\uparrow}^{b}c_{j,\downarrow}^{b^{\dagger}}c_{j,\downarrow}^{b}+c_{j,\uparrow}^{a^{\dagger}}c_{j,\uparrow}^{a}c_{j,\downarrow}^{a^{\dagger}}c_{j,\downarrow}^{a}\right),\\
H_{\Delta}=&&\Delta\sum_{j=1,\sigma}^{L}\left( c_{j,\sigma}^{b^{\dagger}}c_{j,\sigma}^{b}-c_{j,\sigma}^{a^{\dagger}}c_{j,\sigma}^{a}\right).
\end{eqnarray}
\end{subequations}
\endgroup
The parameters $t, U$ and $\Delta$ are hopping integral between two adjacent sublattices, on-site Hubbard interaction and ionic terms, respectively. 
In this paper, we set the energy unit by $t=1$. 
The superscripts $a,b$ are introduced to label the two sublattices with ionic potentials $\pm\Delta$, respectively. 
Therefore $c_{j,\sigma}^{\eta^{\dagger}}$ and $c_{j,\sigma}^{\eta}$ are the creation and annihilation operators for an electron of spin $\sigma$ 
in the Wannier state at the $j$th unit cell as in Fig.~\ref{figure1}
and $\eta=a,b$ labels the internal orbital degree of freedom of every unit cell that corresponds to two sublattices. The Fermionic anticommutation relations is given by 
$\{ c_{j,\sigma}^{\eta^{\dagger}},c_{j',\sigma'}^{\eta'}\} =\delta_{j,j'}\delta_{\eta,\eta'}\delta_{\sigma,\sigma'}$. 
They act in a Fock space with the pseudovacuum $\vert{0}\rangle$ defined by $c_{j,\sigma}^{\eta}\vert{0}\rangle=0$.

The model has translational symmetry in the $j$ and rotational invariance in spin space, with the corresponding SU(2) generators given by,
\begingroup\makeatletter\def\f@size{10.5}\check@mathfonts
\begin{subequations}
\begin{eqnarray}
&&S^{x}=\frac{1}{2}\sum_{j=1}^{L}\sum_{\eta}\left(c_{j,\uparrow}^{\eta^{\dagger}}c_{j,\downarrow}^{\eta}+c_{j,\downarrow}^{\eta^{\dagger}}c_{j,\uparrow}^{\eta}\right),\\
&&S^{y}=\frac{i}{2}\sum_{j=1}^{L}\sum_{\eta}\left( c_{j,\downarrow}^{\eta^{\dagger}}c_{j,\uparrow}^{\eta}-c_{j,\uparrow}^{\eta^{\dagger}}c_{j,\downarrow}^{\eta}\right),\\
&&S^{z}=\frac{1}{2}\sum_{j=1}^{L}\sum_{\eta}\left( c_{j,\uparrow}^{\eta^{\dagger}}c_{j,\uparrow}^{\eta}-c_{j,\downarrow}^{\eta^{\dagger}}c_{j,\downarrow}^{\eta}\right).
\end{eqnarray}
\end{subequations}
\endgroup

The number operators 
$\hat{n}_{\uparrow}=\sum_{j=1}^{L}\sum_{\eta}c_{j,\uparrow}^{\eta^{\dagger}}c_{j,\uparrow}^{\eta}$ and 
$\hat{n}_{\downarrow}=\sum_{j=1}^{L}\sum_{\eta}c_{j,\downarrow}^{\eta^{\dagger}}c_{j,\downarrow}^{\eta}$ 
correspond to $\up$ and $\down$ spins, from which
the total number of electrons will become,
$\hat{n}=\sum_{j=1}^{L}\sum_{\eta, \sigma}c_{j,\sigma}^{\eta^{\dagger}}c_{j,\sigma}^{\eta}$.
Obviously the commutation relations $\left[H,\hat{n}\right]=\left[H,\hat{n}_{\uparrow}\right]=\left[H,\hat{n}_{\downarrow}\right]=0$ hold
that imply the total number of electrons $N$, down-spin electrons, $M$, and up-spin electrons, $(N-M)$, are constants. 
Therefore we can label the eigenstates in sectors specified by $N$ and $M$.

In at attempt to overcome the Choy-Haldane issue~\cite{Choy_1982}, a projection of kinetic energy term into
a subspace with no more than two electrons per unit cell was proposed~\cite{LEE1990398,Frahm_1993}.
In our case it will amount to define a constrained IHM (CIHM) by 
$ H_{\rm CIHM}={\cal P}H_{\rm t}{\cal P}+H_{\rm U}+H_{\Delta}$
where ${\cal P}$ is the projection operator into the subspace with no more than two electrons per unit cell.
This deformation of IHM restricts the Hilbert space in a way that it avoids the Choy-Haldane issue. But 
unfortunately it spoils the $U\to 0$ limit. In fact the $U\to 0$ limit of the Bethe ansatz wave function falls
outside the portion of Hilbert space of the CIHM. However, within the IHM itself, when the Hubbard $U$
is large enough, and the filling is low-enough, the double occupancy of individual orbitals 
will be effectively avoided. Therefore our Bethe wave function obtained in this paper will be
effective description of the spectrum of IHM in this limit.

In the following we will construct a generalization of Bethe wave function
of Lieb and Wu for non-zero ionic potential $\Delta$. We show that its two-particle
scattering matrix satisfies the Yang-Baxter equation. This wave functions allows us to explore strong $U$ limit of the IHM
at low filling.

\section{IHM MOTIVATED BETHE ANSATZ WAVE FUNCTION}
In this section we discuss the coordinate Bethe ansatz wave function inspired by IHM.
In this sense it will be a generalization of Lieb-Wu wave function for nonzero $\Delta$. 
From this equation we construct the two-particle scattering matrix and show that it satisfies the Yang-Baxter equation.
Let the Hilbert space of $N$ electrons be $\mathcal{H}_{N}$ that contains a vacuum state $\vert0\rangle$. 
In this Hilbert space, we search for wave functions of the following Bethe ansatz form,
\begin{eqnarray}
\vert F\rangle=&&\sum_{x_1, ... , x_N}\sum_{\eta_1, ... , \eta_N}\sum_{\sigma_1, ... , \sigma_N}\nonumber\\&&F_{\sigma_1, ... , \sigma_N}(x_1,\eta_1 ; ... ; x_N,\eta_N)\prod_{j=1}^{N}c_{x_j,\sigma_j}^{\eta_j^{\dagger}}\vert0\rangle,
\end{eqnarray}
where $j=1, \ldots ,N$ labels the electrons, $x_j$ and $\sigma_j$ are position and spin of $j$'th electron, $\eta_j=a,b$ is the sublattice label of the $j$'th electron and
\begingroup\makeatletter\def\f@size{9.0}\check@mathfonts
$$\textcolor{red}{F_{\bsSigma}(\bsX,\bseta)\equiv F_{\sigma_1\ldots\sigma_N}(x_1,\eta_1;\ldots ;x_N,\eta_N)=\mathcal{A}\prod_{j=1}^{N}e^{ik_jx_j}\chi_{\eta_j}(\theta_j)\sum_{Q}A_{\sigma_1, ... , \sigma_N}(Q)\tilde{\mathbb{H}}(x_Q)}
$$
\endgroup
is the BA wave function in coordinate basis \hlt{the various components of which are further specified in the following}. 
The Schr\"odinger equation $H\vert F\rangle =E\vert F\rangle$ in the coordinate basis becomes $h_{N} F_\bsSigma(\bsX,\bseta) =E F_\bsSigma(\bsX,\bseta)$ 
with the first quantized Hamiltonian $h_N$ given by,
\begingroup\makeatletter\def\f@size{10.5}\check@mathfonts
\bearr
h_{N}&=\sum_{j=1}^{N}\left( -2t\cos\left(\frac{k_j}{2}\right)\left(\cos\left(\frac{k_j}{2}\right)\tau_{j}^{x}+\sin\left(\frac{k_j}{2}\right)\tau_{j}^{y}\right)+\Delta\tau_{j}^{z}\right) \nn\\
&+\frac{U}{2}\sum_{j\leq l}\delta_{x_j,x_l}\left(\tau_{j}^{0}\otimes\tau_{l}^{0}+\tau_{j}^{z}\otimes\tau_{l}^{z}\right),
\eearr
\endgroup
where $\tau_{j}^{i}$ is the $i$th ($i=0,x,y,z$) unit and Pauli matrix acting in sublattice space $a,b$ of the $j$'th electron.

In the one-particle sector ($N= 1$) the Schr\"odinger equation becomes, 
\begingroup\makeatletter\def\f@size{9.0}\check@mathfonts
\begin{equation}
\left[-2t \cos(\frac{k}{2})\left( \cos(\frac{k}{2})\tau^{x}+\sin(\frac{k}{2})\tau^{y}\right)+\Delta\tau^{z}\right] F_{\sigma}^{\eta}(x)=E F_{\sigma}^{\eta}(x),\nn
\end{equation} 
\endgroup
whose solution are,
\begingroup\makeatletter\def\f@size{9.5}\check@mathfonts
\begin{equation}\label{N1sol}
\begin{split}
&F_{\sigma}^{\eta}(x)=A_{\sigma}e^{ikx}\chi_{\eta}(\theta),\; \\&E(\theta)=\Delta\cosh(\theta),\; \\&k(\theta)=2\cos^{-1}(\frac{\Delta}{2t}\sinh(\theta)).
\end{split}
\end{equation} 
\endgroup
where $\theta$ is the rapidity of the particle, $k$ is the wave vector, $A_{\sigma}$ is a spinor in spin space and 
$\chi_{\eta}(\theta)$ is a spinor in sublattice space whose components are of the following form,
\begingroup\makeatletter\def\f@size{10.5}\check@mathfonts
\begin{equation}
\chi_{b}(\theta)=-\varkappa(\theta)\frac{\cosh\left(\frac{\theta}{2} \right) }{\sqrt{\cosh(\theta)}},\quad \chi_{a}(\theta)=
\frac{\sinh\left(\frac{\theta}{2} \right) }{\sqrt{\cosh(\theta)}}.
\end{equation} 
\endgroup
In this equation, $\varkappa(\theta)=\frac{\Delta}{2t}\sinh(\theta)+i\sqrt{\frac{4t^2}{\Delta^2}-\sinh^2 (\theta)}$.
There are two bands with positive and negative energies. Eq.~\eqref{N1sol} is for positive energy states. The negative
energy states are simply obtained by $\theta\to i\pi-\theta$. 

In construction of BA for models (such as Hubbard model) with only one orbital degree of freedom per unit cell, one breaks up
the region of spatial variables $(x_1, \ldots, x_N )$ into $N!$ corners obtained by operating permutation $Q$ of indices 
$1,2\ldots,N$ on $x_1<x_2\ldots<x_N$ and looking for a wavefunction in each of these corners. 
For $N=2$ there are $N!=2$ corners, namely $x_1<x_2$ and $x_2<x_1$. 
In the case of IHM where there is additional orbital degree of freedom labelled by $\eta=a,b$, we need to slightly adjust the notation
to take proper care of the spatial ordering of electron coordinates. In this case the spatial coordinates are $(x_1,\eta_1 ; ... ; x_N,\eta_N)$
and a modified step function will be introduced to handle the ordering of the actual location of electrons. 
Let us see how this is done for $N=2$ electrons. Higher $N$ values is similarly treated. 
The Hamiltonian for two electrons is $h_{2}=\sum_{j=1}^{2}\left( -2t\cos\left(\frac{k_j}{2}\right)\left(\cos\left(\frac{k_j}{2}\right)\tau_{j}^{x}+\sin\left(\frac{k_j}{2}\right)\tau_{j}^{y}\right)+\Delta\tau_{j}^{z}\right)+\frac{U}{2}\delta_{x_1,x_2}\left(\tau_{1}^{0}\otimes\tau_{2}^{0}+\tau_{1}^{z}\otimes\tau_{2}^{z}\right)$. 
The Hubbard interaction $U$ operates when $x_1=x_2$ (i.e. the particles are in the same unit cell), and further $\eta_1=\eta_2$ meaning that they are
on the same sublattice. Away from the boundary of the corners the Hamiltonian is free and the wavefunction can be written in the form of product of single particle solutions,
\begingroup\makeatletter\def\f@size{10.5}\check@mathfonts
\bearr
&&F_{\sigma_1, \sigma_2}(x_1,\eta_1 ; x_2,\eta_2)= \nn\\
&&\left(A_{\sigma_1, \sigma_2}\tilde{\mathbb{H}}(x_1- x_2)+B_{\sigma_1, \sigma_2}\tilde{\mathbb{H}}(x_2- x_1)\right)\nn\\
&&\times e^{i(k_1x_1+k_2x_2)}\chi_{\eta_1}(\theta_1)\chi_{\eta_2}(\theta_2)\nn\\
&&-\left(A_{\sigma_2, \sigma_1}\tilde{\mathbb{H}}(x_2- x_1)+B_{\sigma_2, \sigma_1}\tilde{\mathbb{H}}(x_1- x_2)\right)\nn\\
&&\times e^{i(k_1x_2+k_2x_1)}\chi_{\eta_1}(\theta_2)\chi_{\eta_2}(\theta_1),
\label{N2WF}
\eearr
\endgroup
Here
$\tilde{\mathbb{H}}(x_j- x_l)=\mathbb{H}(x_j- x_l)+\frac{\delta_{x_j,x_l}}{2}\left(\delta_{\eta_j,b}\delta_{\eta_l,a}-\delta_{\eta_j,a}\delta_{\eta_l,b}\right)$ 
is the extended Heaviside step function
that separates the two corners by incorporating their orbital degree of freedom $\eta=a,b$ of the $j$'th and $l$'th electrons. 
This wave function is explicitly antisymmetric. 
\hlt{Furthermore, $\mathbb{H}$ is the ordinary Heaviside step function and is regularized as by defining $\mathbb{H}(0)=\frac{1}{2}$}. 
The total energy and total momentum for this wave function are given by 
$E=E_1+E_2=\Delta\left(\cosh(\theta_1)+\cosh(\theta_1)\right)$ and 
$k=k_1+k_2=2\left( \cos^{-1}(\frac{\Delta}{2t}\sinh(\theta_1))+\cos^{-1}(\frac{\Delta}{2t}\sinh(\theta_2))\right)$, respectively, that are conserved in all corners.

We introduce scattering matrix relating to the spin amplitudes in two corners of Eq.~\eqref{N2WF} as
$B_{\sigma_1, \sigma_2}=S_{\sigma_1, \sigma_2}^{\sigma'_1, \sigma'_2}A_{\sigma'_1, \sigma'_2}$. This is determined by imposing two conditions~\cite{andrei1992integrable}:\\
($i$) \textit{Continuity of the wavefunction in the corner boundary (Uniqueness)}. 
Considering the condition of the continuity of the wavefunction at $x_1=x_2$ and $\eta_1=\eta_2$ leads to,
\begingroup\makeatletter\def\f@size{10.5}\check@mathfonts
\begin{equation}\label{Ssigma}
\textbf{S}=\frac{1}{2}\left(\textbf{I}+\textbf{P}\right)+\frac{1}{2}\left(\textbf{I}-\textbf{P}\right)\textbf{\textit{s}},
\end{equation} 
\endgroup
where $\textbf{I}_{\sigma_1,\sigma_2}^{\sigma'_1,\sigma'_2}=\delta^{\sigma'_1}_{\sigma_1}\delta^{\sigma'_2}_{\sigma_2}$ and 
$\textbf{P}_{\sigma_1,\sigma_2}^{\sigma'_1,\sigma'_2}=\delta^{\sigma'_2}_{\sigma_1}\delta^{\sigma'_1}_{\sigma_2}$ are the identity and exchange operators in the
spin space, respectively. In the ordinary Hubbard model where there is one orbital degree of freedom per unit cell, $\textbf{\textit s}$ has to be a scalar (which is
called spin scalar parameter in the BA literature).
But in the present case where we have two orbitals (sublattices) $a,b$ in each unit cell, $\textbf{\textit s}$ will be a matrix in the orbital space that
completely characterizes the scattering process. \\
($ii$) \textit{The Schr\"odinger equation on the corner boundary} ($x_1=x_2; \eta_1=\eta_2$). 
The Schr\"odinger equation for two electrons has four components labelled with sublattice indices $\textbf{bb}, \textbf{ba}, \textbf{ab}$ and $\textbf{aa}$. 
For two of the four cases, namely, ($\textbf{bb}, \textbf{aa}$) Hubbard interactions occur between electrons. Therefore $\textbf{\textit s}$ 
for these components are obtained as follows (for details see appendix A),
\begingroup\makeatletter\def\f@size{10.5}\check@mathfonts
\begin{equation}\label{ssp-bb-aa}
\textit{\textbf{s}}^{\textbf{b}\textbf{b}}=\frac{\mathfrak{\Xi}^{\textbf{b}\textbf{b}}-\frac{U}{t}}{\mathfrak{\Xi}^{\textbf{b}\textbf{b}}+\frac{U}{t}},\qquad \textit{\textbf{s}}^{\textbf{a}\textbf{a}}=\frac{\mathfrak{\Xi}^{\textbf{a}\textbf{a}}-\frac{U}{t}}{\mathfrak{\Xi}^{\textbf{a}\textbf{a}}+\frac{U}{t}},
\end{equation} 
\endgroup
where rapidity related parts are $\mathfrak{\Xi}^{\textbf{b}\textbf{b}}=\left(\frac{\chi_{a}(\theta_2)}{\chi_{b}(\theta_2)}\left(1-e^{ik_2}\right)-\frac{\chi_{a}(\theta_1)}{\chi_{b}(\theta_1)}\left(1-e^{ik_1}\right)\right)$,
and 
$\mathfrak{\Xi}^{\textbf{a}\textbf{a}}=\left(\frac{\chi_{b}(\theta_1)}{\chi_{a}(\theta_1)}\left(1-e^{-ik_1}\right)-\frac{\chi_{b}(\theta_2)}{\chi_{a}(\theta_2)}\left(1-e^{-ik_2}\right)\right)$. 
It can be seen that, by changing the sign of $\Delta$, these two amplitudes transform to each other, namely,
$\textit{\textbf{s}}^{\textbf{b}\textbf{b}}(\Delta)=\textit{\textbf{s}}^{\textbf{a}\textbf{a}}(-\Delta)$ as expected.
In the limit ($\Delta\rightarrow0$) the scalar spin parameters in Eq.~\eqref{ssp-bb-aa} 
reduce to, 
\begingroup\makeatletter\def\f@size{10.5}\check@mathfonts
\begin{equation}
\begin{split}
\textit{\textbf{s}}^{\textbf{b}\textbf{b}}=\textit{\textbf{s}}^{\textbf{a}\textbf{a}}=\frac{\left(\sin\left(\frac{k_1}{2}\right)-\sin\left(\frac{k_2}{2}\right)\right)-i\frac{U}{2t}}{\left(\sin\left(\frac{k_1}{2}\right)-\sin\left(\frac{k_2}{2}\right)\right)+i\frac{U}{2t}},
\end{split}
\end{equation} 
\endgroup
which has the expected form for the ordinary Hubbard model, except that the argument of the $\sin$ function is $k_j/2$ rather than $k_j$. 
This is because in the present formulation each unit cell contains two degrees of freedom corresponding to which the Brillouin zone will be
folded into half of the case where only one degree of freedom in each unit cell is considered. 
For the other two cases ($\textbf{ba}, \textbf{ab}$), the Hubbard interaction does not appear in their equations.
In this case the parameter $\textbf{\textit s}$ can be written as 
$\textit{\textbf{s}}^{\textbf{b}\textbf{a}}=\textit{\textbf{s}}^{\textbf{a}\textbf{b}}=\zeta-\xi$, with the constraint that $\zeta+\xi=1$. 
This is because when two electrons do not interact with each other, they can either pass each other without change (scattering matrix is unit matrix $\textbf{\textit I}$),
or exchange (scattering matrix is solely $\textbf{\textit P}$~\cite{andrei1992integrable}). Any linear combination of the above two cases is also acceptable and this
is the source of degree of arbitrariness in choosing $\zeta$ and $\xi$. The only constraint is $\zeta+\xi=1$ which stems from the solution of Schr\"odinger
equation at the boundary. 
By choosing $\zeta=0$ and $\xi=1$, the spin scalar parameter will be 
$\textit{\textbf{s}}^{\textbf{b}\textbf{a}}=\textit{\textbf{s}}^{\textbf{a}\textbf{b}}=-1$ 
and we can write them as follows:
\begingroup\makeatletter\def\f@size{10.5}\check@mathfonts
\begin{equation}\label{ssp-ba-ab}
\textit{\textbf{s}}^{\textbf{b}\textbf{a}}=\textit{\textbf{s}}^{\textbf{a}\textbf{b}}=\frac{0-\frac{U}{t}}{0+\frac{U}{t}}.
\end{equation} 
\endgroup

Finally, by combining various components of the spin scalar in Eqs.~(\ref{ssp-bb-aa}) and~(\ref{ssp-ba-ab}), 
and by taking into account the probability of each $(a,b)$ components of the electrons, the spin scalar parameter becomes,
\begingroup\makeatletter\def\f@size{10.5}\check@mathfonts
\begin{equation}\label{ssp-t}
\begin{split}
\textit{\textbf{s}}=&\left(\vert\chi_{b}(\theta_1)\vert^2\vert\chi_{b}(\theta_2)\vert^2\mathfrak{\Xi}^{\textbf{b}\textbf{b}}+\vert\chi_{b}(\theta_1)\vert^2\vert\chi_{a}(\theta_2)\vert^2\times0\right.\\
&\left.+\vert\chi_{a}(\theta_1)\vert^2\vert\chi_{b}(\theta_2)\vert^2\times0+\vert\chi_{a}(\theta_1)\vert^2\vert\chi_{a}(\theta_2)\vert^2\mathfrak{\Xi}^{\textbf{a}\textbf{a}}-
U/t\right)\\&/\left(\vert\chi_{b}(\theta_1)\vert^2\vert\chi_{b}(\theta_2)\vert^2\mathfrak{\Xi}^{\textbf{b}\textbf{b}}+\vert\chi_{b}(\theta_1)\vert^2\vert\chi_{a}(\theta_2)\vert^2\times0\right.\\
&\left.+\vert\chi_{a}(\theta_1)\vert^2\vert\chi_{b}(\theta_2)\vert^2\times0+\vert\chi_{a}(\theta_1)\vert^2\vert\chi_{a}(\theta_2)\vert^2\mathfrak{\Xi}^{\textbf{a}\textbf{a}}+
U/t\right).
\end{split}
\end{equation} 
\endgroup
By inserting Eq.~(\ref{ssp-t}) into Eq.~(\ref{Ssigma}) and after some algebra, and using the 
definition $\alpha(\theta)\equiv\tanh(\theta)\sqrt{4t^2/\Delta^2-\sinh^2(\theta)}$, the two particle scattering matrix will be,
\begingroup\makeatletter\def\f@size{10.5}\check@mathfonts
\begin{equation}\label{Sthetat}
\begin{split}
\textbf{S}_{i,j}=&\frac{\left(\alpha(\theta_i)-\alpha(\theta_j)\right)\textbf{I}_{i,j}+i\frac{2U}{\Delta}\textbf{P}_{i,j}}{\left(\alpha(\theta_i)-\alpha(\theta_j)\right)+i\frac{2U}{\Delta}}.
\end{split}
\end{equation} 
\endgroup
By changing the variable from the rapidity $\theta$ to momentum $k$ according to Eq.~\eqref{N1sol}, 
for particles with positive energies we have $\alpha(k)\equiv\frac{2t\cos(k/2)\sin(k/2)}{\sqrt{\Delta^2+4t^2\cos^2(k/2)}}$ which 
then gives,
\begingroup\makeatletter\def\f@size{10.5}\check@mathfonts
\begin{equation}\label{Skt}
\begin{split}
\textbf{S}_{i,j}=&\frac{\left(\alpha(k_i)-\alpha(k_j)\right)\textbf{I}_{i,j}+i\frac{U}{t}\textbf{P}_{i,j}}{\left(\alpha(k_i)-\alpha(k_j)\right)+i\frac{U}{t}}.
\end{split}
\end{equation} 
\endgroup
From Eq.~(\ref{Skt}) it is clear that the limit of $\Delta\to 0$, this scattering matrix is exactly the same as the scattering matrix obtained for the 
standard Hubbard model. The only difference is that $U$ has been replaced with $2U$. 
This is because the present scattering matrix is for a unit cell containing an orbital degree of freedom $\eta=a,b$. 
Thus, each electron in interaction with another electron is experiencing a total potential of $2U$.

A straightforward calculation shows that the above $S$ matrix obtained for the IHM satisfies the Yang-Baxter equation (for details see
appendix B). Therefore the following BA wavefunction for the $N$-particle system
is consistent at three-particle scattering level and therefore {\it it describes the exact wave function of 
some integrable model},
\begin{eqnarray}
&&F_{\sigma_1, ... , \sigma_N}(x_1,\eta_1 ; ... ; x_N,\eta_N) \label{NPWF} \\
&&=\mathcal{A}\prod_{j=1}^{N}e^{ik_jx_j}\chi_{\eta_j}(\theta_j)\sum_{Q}S(Q)A_{\sigma_1, ... , \sigma_N}\tilde{\mathbb{H}}(x_Q).\nn
\end{eqnarray}
Here $Q$ is a corner corresponding to one set of permutations, $S(Q)$ is the corresponding product of $S$-matrices, 
and $A_{\sigma_1, ... , \sigma_N}$ is the spin amplitude in some reference corner. The extended Heaviside function
is defined below Eq.~\eqref{N2WF} which incorporates the sublattice structure into the definition of corners. 
The large $U$ and low-filling limit of the model described by the above wave function, approximately corresponds to the IHM. 

Now we would like to study the system of $0<N\leqslant L$ electrons on a finite length $L$ with periodic boundary conditions. 
We can impose this condition in different regions in configuration space to the wavefunction as,
\begingroup\makeatletter\def\f@size{10.5}\check@mathfonts
\begin{equation}
\begin{split}
F_{\sigma_1, ... , \sigma_N}&(... ;x_j=1,\eta_j ; ...)\\&=F_{\sigma_1, ... , \sigma_N}(... ;x_j=L+1,\eta_j ; ...).
\end{split}
\end{equation} 
\endgroup
For the configuration $Q$($x_1<x_2< ... <x_N$) the above relation is translated to conditions on the spin amplitude of wavefunction 
in a single corner as,
\begingroup\makeatletter\def\f@size{10.5}\check@mathfonts
\begin{equation}
\begin{split}
(Z_j)_{\sigma_1, ... , \sigma_N}^{\sigma'_1, ... , \sigma'_N}A_{\sigma'_1, ... , \sigma'_N}(Q)=e^{-ik_jL}A_{\sigma_1, ... , \sigma_N}(Q),
\end{split}
\end{equation} 
\endgroup
where \hlt{$Z_j=S^{j,j-1}S^{j,j-2} \ldots S^{j,1}S^{j,N}S^{j,N-1} \ldots S^{j,j+2}S^{j,j+1}$}, is the so called the transfer matrix. Again, we are encountered 
with a eigenvalue problem of diagonalization of the spin Hamiltonian ($Z_j$) by using of the quantum inverse scattering method for a state 
with $M$ down spins and $N-M$ up spins and obtain Bethe Ansatz equations as follows~\cite{PhysRevLett.21.192.2,andrei1992integrable},
\begingroup\makeatletter\def\f@size{10.5}\check@mathfonts
\begin{subequations}\label{BAE}
\begin{equation}
\begin{split}
e^{ik_jL}=\prod_{\delta=1}^{M}\frac{\Lambda_\delta-\alpha_j-ic/2}{\Lambda_\delta-\alpha_j+ic/2},\qquad j=1,2, ... , N
\end{split}
\end{equation} 
\begin{equation}
\begin{split}
-\prod_{\delta=1,\delta\neq\gamma}^{M}\frac{\Lambda_\delta-\Lambda_\gamma+ic}{\Lambda_\delta-\Lambda_\gamma-ic}=\prod_{j=1}^{N}\frac{\Lambda_\gamma-\alpha_j-ic/2}{\Lambda_\gamma-\alpha_j+ic/2},\; \gamma=1,2, ... ,M
\end{split}
\end{equation} 
\end{subequations}
\endgroup
where $c=\frac{2U}{\Delta}$, $\alpha_j\equiv\alpha(\theta_j)$ is defined in Eq.~(\ref{Sthetat}) and $M$ labels the number of down spins in a specific configuration.
The spin variables $\Lambda_\gamma$ are coupled with momenta $k_j$ and must be determined self-consistently as detailed
in the next section. 

\section{THERMODYNAMIC LIMIT AND GROUND STATE PROPERTIES}
As already mentioned, we are going to solve these equations for the $0<N\leqslant L$ where $N$ is the total number of electrons
and $L$ is the number of {\em unit cells}. We will consider the filling factors ($n=N/L\leqslant1$) of less than a quarter filled bands. 
This amounts to avoiding the Choy-Haldane issue {\em on average} which makes our solution relevant to the IHM. 
From here on, we write equations in the wave vector space instead of rapidities. At low enough temperatures, only the lower band
will be relevant and therefore, the $\alpha_j$ parameter in Eqs.~(\ref{BAE}) will be in the form of 
$\alpha_j=\frac{2t\cos(k_j/2)\sin(k_j/2)}{\sqrt{\Delta^2+4t^2\cos^2(k_j/2)}}$ where $c=U/t$. We will be 
interested in the limit $N,M,L\to \infty$ with $n=N/L$ and $M/L$ held constant. 

For a finite system with $N$ electrons the total energy and momentum of the system given by,
\begingroup\makeatletter\def\f@size{10.5}\check@mathfonts
\begin{equation}\label{En.and.Mom}
\begin{split}
E=-\sum_{j=1}^{N}\sqrt{\Delta^2+4t^2\cos^2\left(\frac{k_j}{2}\right)}, \quad P=\sum_{j=1}^{N}k_j.
\end{split}
\end{equation} 
\endgroup
Eigenvalues and eigenfunctions of the system are characterized by set of momenta and spin variables 
$\left\lbrace k_j, \Lambda_\gamma \right\rbrace$ as follows: Taking the logarithm of Eqs.~(\ref{BAE}) we obtain,
\begingroup\makeatletter\def\f@size{10.5}\check@mathfonts
\begin{subequations}\label{LBAE}
\begin{equation}
Lk_j=2\pi I_j+\sum_{\delta=1}^{M}\Theta(2\alpha_j-2\Lambda_\delta),\quad j=1,2, ... , N
\end{equation} 
and
\begin{equation}
\sum_{j=1}^{N}\Theta(2\alpha_j-2\Lambda_\gamma)=2\pi I_\gamma +\sum_{\delta=1}^{M}\Theta(\Lambda_\delta-\Lambda_\gamma) ,\; \gamma=1,2, ... , M
\end{equation} 
\end{subequations}
\endgroup
where $\Theta(x)=-2\tan^{-1}(x/c)$ with condition $-\pi<\Theta(x)<\pi$. The $\left\lbrace I_j,I_\gamma\right\rbrace $ are 
integer or half-integer quantum numbers which labels the states. Using Eqs.~(\ref{LBAE}) the momentum of the system becomes, 
the $P=\frac{2\pi}{L}\left(\sum_{j}I_j +\sum_{\gamma}I_\gamma\right) $. Solutions of Eqs.~(\ref{BAE}) in the thermodynamic limit, 
based on string hypothesis~\cite{takahashi1972one,andrei1992integrable}, can have different forms, one of which is the real $k_j$'s and real $\Lambda_\gamma$'s. 
This is the case for the ground state of the system.

\hlt{According to the Lieb-Mattis theorem~\cite{lieb1962ordering} the ground state of the system on a bipartite lattice should be a total singlet. 
Therefore in our one-dimensional case, the ground state configuration will be specified by the consecutive arrangement of $\left\lbrace I_j\right\rbrace $ 
and $\left\lbrace I_\gamma\right\rbrace $ around zero.}
In the thermodynamic limit \hlt{and for any consecutive arrangements of $\left\lbrace I_j\right\rbrace $ and $\left\lbrace I_\gamma\right\rbrace $ around zero}, we define the densities of solutions $\rho(k)$ and $\sigma(\Lambda)$ by~\cite{Yupeng_Wang_2015},
\begingroup\makeatletter\def\f@size{10.5}\check@mathfonts
\begin{equation}\label{sol.den}
\begin{split}
\rho(k_j)=\frac{1}{L(k_{j+1}-k_j)},\quad \sigma(\Lambda_{\gamma})=\frac{1}{L(\Lambda_{\gamma+1}-\Lambda_{\gamma})}.
\end{split}
\end{equation} 
\endgroup
The values of $\rho(k_j)$ and $\sigma(\Lambda\gamma)$ are continuously distributed in intervals $\left[-Q,Q\right]$ 
and $\left[-B,B\right]$, respectively. The integration limits $Q\leqslant\pi$ and $B\leqslant\infty$ are
specified by the following two conditions:
\begingroup\makeatletter\def\f@size{10.5}\check@mathfonts
\begin{equation}\label{k.qeyd}
\begin{split}
\int_{-Q}^{Q}\rho(k) dk=\frac{N}{L},\qquad \qquad \int_{-B}^{B}\sigma(\Lambda) d\Lambda=\frac{M}{L}.
\end{split}
\end{equation} 
\endgroup

We use the subscript $g$ in the following to indicate that the quantity under consideration is associated with the ground state
of the system. 

Algebraic manipulation of Eqs.~(\ref{LBAE}) for solution densities at the ground state, gives two coupled integral equations~\cite{andrei1992integrable}
\begingroup\makeatletter\def\f@size{9.5}\check@mathfonts
\begin{subequations}\label{ro.sig.integ}
\begin{eqnarray}
\rho_g(k)&=&\frac{1}{2\pi}+f(k)\int_{-B_g}^{B_g}d\Lambda \sigma_g(\Lambda)K_1(\alpha(k)-\Lambda),\\
\sigma_g(\Lambda)&=&\int_{-Q_g}^{Q_g}dk \rho_g(k) K_1(\alpha(k)-\Lambda) \nn\\
&&- \int_{-B_g}^{B_g}d\Lambda' \sigma_g(\Lambda')K_2(\Lambda-\Lambda'),
\end{eqnarray}
\end{subequations}
\endgroup
where $f(k)=\frac{d\alpha(k)}{dk}$, $K_n(x)=\frac{1}{\pi}\frac{(nc/2)}{(nc/2)^2+x^2}$ and its Fourier transform is 
$K_n(\omega)=\frac{1}{\sqrt{2\pi}}e^{-n\frac{c}{2}\vert\omega\vert}$. For the ground state, 
the $z$-component and total spin is equal to zero ($S_z=S=\frac{1}{2}(N-2M)$) such that one has $B_g=\infty$. Then by using the Fourier transform 
of $\sigma_g(\Lambda)=1/\sqrt{2\pi}\int_{-\infty}^{\infty}d\omega \sigma_g(\omega) e^{i\Lambda\omega} $ for $\rho_g(k)$ we have,
\begingroup\makeatletter\def\f@size{10.5}\check@mathfonts
\begin{equation}\label{ro.g.k}
\begin{split}
\rho_g(k)=\frac{1}{2\pi}+\frac{2}{c}f(k)\int_{-Q_g}^{Q_g}dk' \rho_g(k')R\left(\frac{2}{c}\left(\alpha(k)-\alpha(k')\right)\right).
\end{split}
\end{equation} 
\endgroup
In this equation, $R(x)\equiv(1/4\pi)\int_{-\infty}^{\infty}dy \frac{e^{i \frac{xy}{2}}}{1+e^{|y|}}=(1/\pi)\sum_{n=1}^{\infty}(-1)^{n+1}\frac{2n}{x^2+(2n)^2}$ 
is a real valued function. Thus the ground state energy can be written as,
\begingroup\makeatletter\def\f@size{10.5}\check@mathfonts
\begin{equation}\label{gs.energy}
\begin{split}
E_g(k)=-L\int_{-Q_g}^{Q_g}dk \rho_g(k)\sqrt{\Delta^2+4t^2\cos^2\left(\frac{k}{2}\right)}.
\end{split}
\end{equation} 
\endgroup
For quarter filling ($N/L=1$) one has $Q_g=\pi$, and the density of electron states ($\rho_g(k)$) and ground state energy can be 
analytically calculated as,
\begingroup\makeatletter\def\f@size{10.5}\check@mathfonts
\begin{subequations}\label{gs.quarter.energy}
\begin{equation}
\begin{split}
\rho_g(k)=\frac{1}{2\pi}+\frac{f(k)}{\pi}\int_{0}^{\infty}d\omega \cos(\omega\alpha(k))\frac{\mathcal{J}_0(\omega)}{1+e^{c\omega}},
\end{split}
\end{equation} 
\begin{equation}
\begin{split}
E_g(k)=-2L\left(\frac{\Delta}{\pi}E\left[-\frac{4t^2}{\Delta^2}\right]+\int_{0}^{\infty}d\omega \frac{\mathcal{J}_0(\omega)\mathcal{J}_1(\omega)}{\omega\left( 1+e^{c\omega}\right)}\right),
\end{split}
\end{equation} 
\end{subequations}
\endgroup
where $E(\nu)=\int_{0}^{\pi/2}dx\sqrt{1-\nu\sin^2(x)}$ , is the complete elliptic integral of the second kind.
The functions $\mathcal{J}_0$ and $\mathcal{J}_1$ are defined by $\mathcal{J}_0(\omega)\equiv\pi^{-1}\int_{0}^{\pi}dk \cos(\omega\alpha(k))$, 
and $\mathcal{J}_1(\omega)\equiv\omega\pi^{-1}\int_{0}^{\pi}dk \cos(\omega\alpha(k))f(k)\sqrt{\Delta^2+4t^2\cos^2\left(\frac{k}{2}\right)}$.
In the limit $\Delta\to 0$, these functions reduce to the standard Bessel functions $J_0$ and $J_1$ of the Lieb-Wu solution~\cite{lieb1994absence}. 
Therefore our expression for the ground state energy in the limit of $\Delta\to 0$, correctly reduces to the Lieb-Wu result. 

\begin{figure}[t]
\includegraphics[scale=0.18]{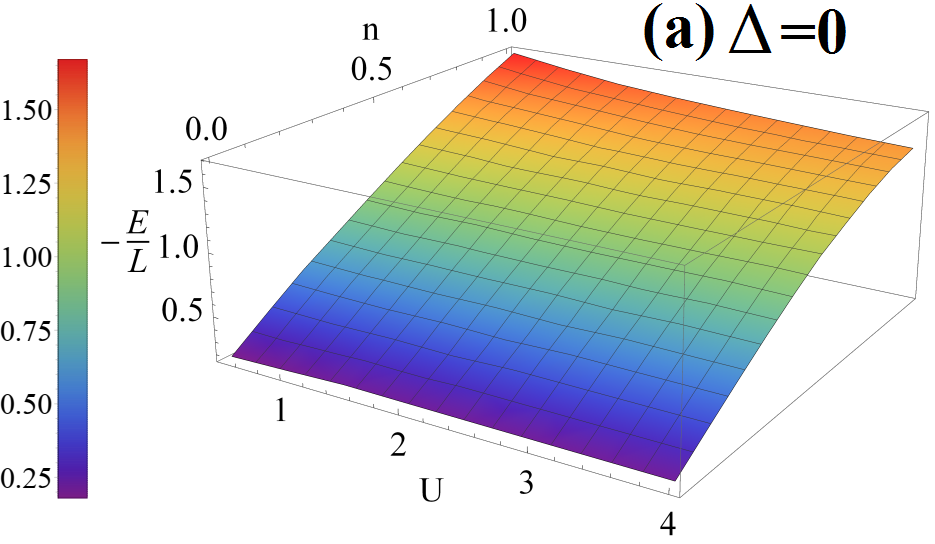}\hfill
\includegraphics[scale=0.10]{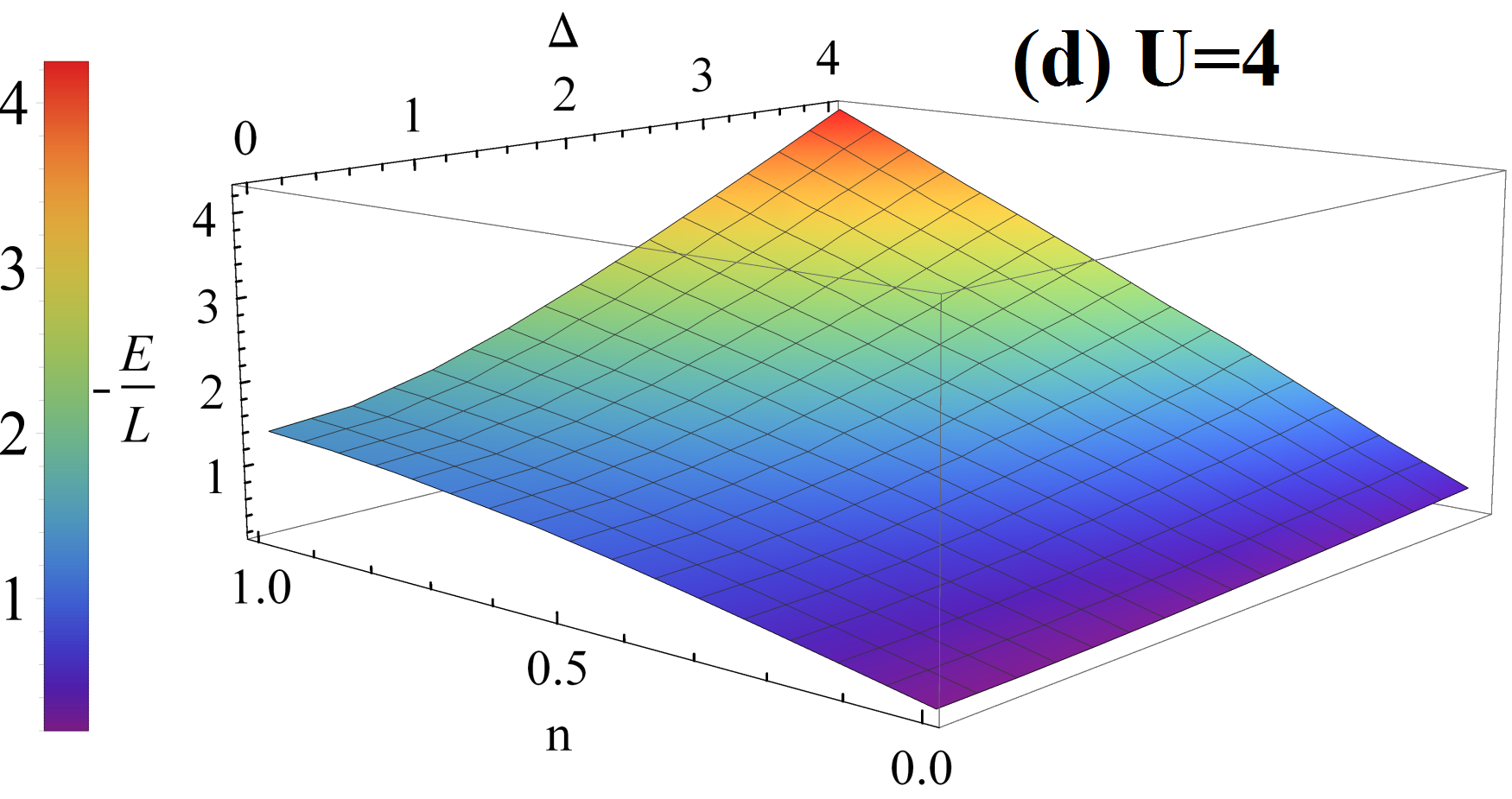}
\includegraphics[scale=0.35]{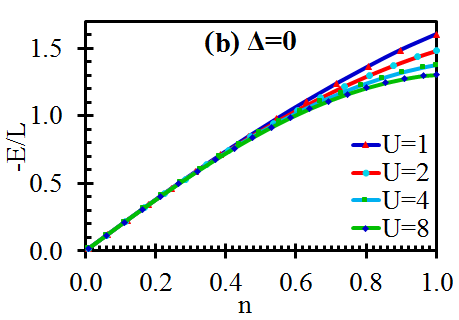}\hfill
\includegraphics[scale=0.24]{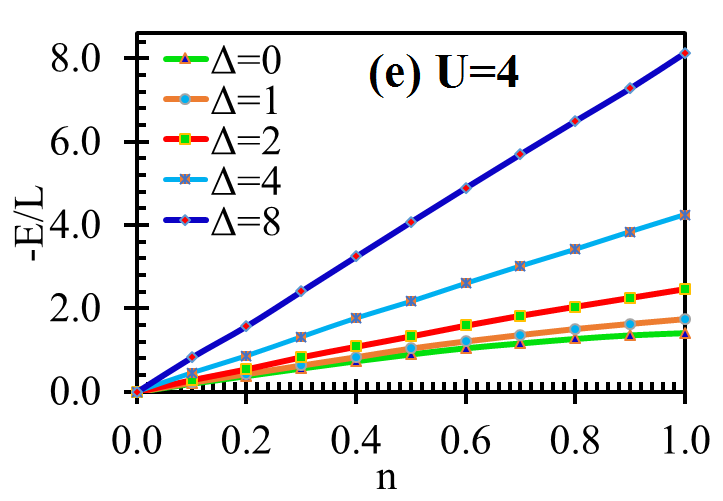}
\includegraphics[scale=0.24]{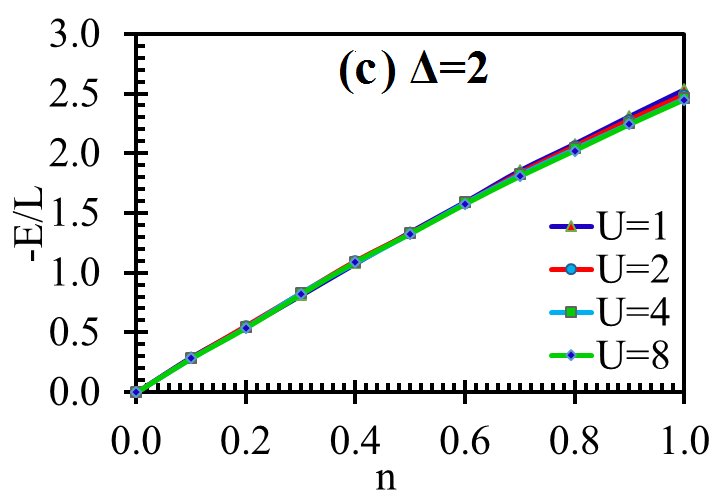}\hfill
\includegraphics[scale=0.24]{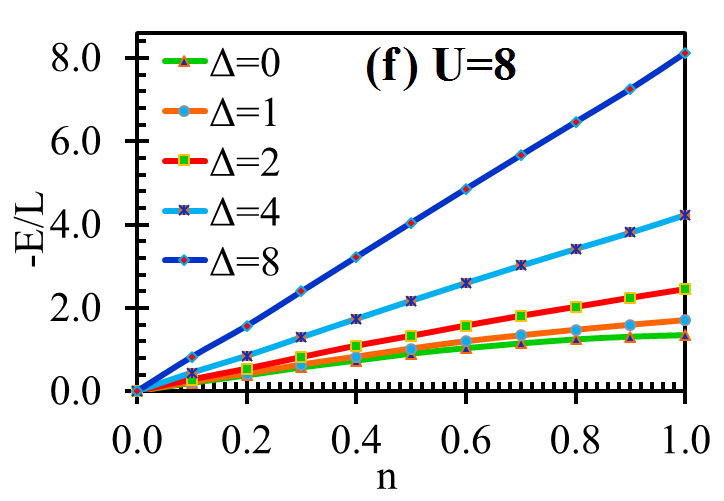}
\caption{\label{figure4} The ground state energy of our Bethe ansatz wave function as a function of $U, \Delta$ and $n$. The unit of energy is taken to be $T=1$. 
The dependence of ground state energy to electron density and Hubbard parameter $U$
for ionic potential (a) $\Delta=0$ (the standard Hubbard model). Panel (b) is represents constant $U$ cuts of Panel (a). 
Panel (c) is similar to (b), except that now $\Delta=2.0$. 
In the second column we represent the ground state energy as a function of $\Delta$ and $n$ for fixed (d) $U=4.0$,
(e) constant $\Delta$ cuts from panel (d). Panel (f) is the same constant $\Delta$ plots for strong Hubbard $U=8.0$. }
\end{figure}

To obtain the ground state energy for different conditions, we numerically solve the equations~(\ref{k.qeyd}), (\ref{ro.g.k}) and (\ref{gs.energy}). 
Fig.~\ref{figure4} shows our results for ground-state energy. 
In panels (a) and (b) we plot the ground state energy as a function of $U$ and $n$ for $\Delta=0$ that corresponds to the standard Hubbard model~\cite{shiba1972magnetic}. 
In panel (c) we increase the ionic potential to $\Delta=2.0$. In the second column we plot the ground state energy as a function of $\Delta$ and $n$ 
for a fixed $U$. The values of $U$ in panels (d) and (e) are $U=4.0$ while in panel (f) we have a quite strong Hubbard $U=8.0$. 
First of all, the $\Delta=0$ results in (a) and (b) agree with the numerical results for the Hubbard model~\cite{shiba1972magnetic}.
As can be seen in panel (b), at low densities the chemical potential (slope of the curve) is not much dependent on $U$. But
by increasing $n$ towards quarter filling, the $U$ dependence of the slope (chemical potential) becomes important. This is in contrast
to panel (c) with $\Delta=2.0$ where even near the quarter filling the slope is not much sensitive to the values of $U$. 
Near the quarter-filling ($n=1$) there is in average one electrons in a unit cell which occupies the negative $\Delta$ site. 
Therefore for large enough $\Delta$, one barely faces a situation with two electrons coming across each other at the same site
and that is why the $U$ dependence upto quarter filling is not strong.

Comparing panels (e) and (f) corresponding to $U=4.0$ and $U=8.0$, respectively,
it can be seen that the main factor controlling the shape of the curves is $\Delta$, and the
$U$-dependence is already very weak. It appears that the large $U$ regime of validity of our wave function
already extends down to $U\sim 4$.

\begin{figure}[t]
\includegraphics[scale=0.205]{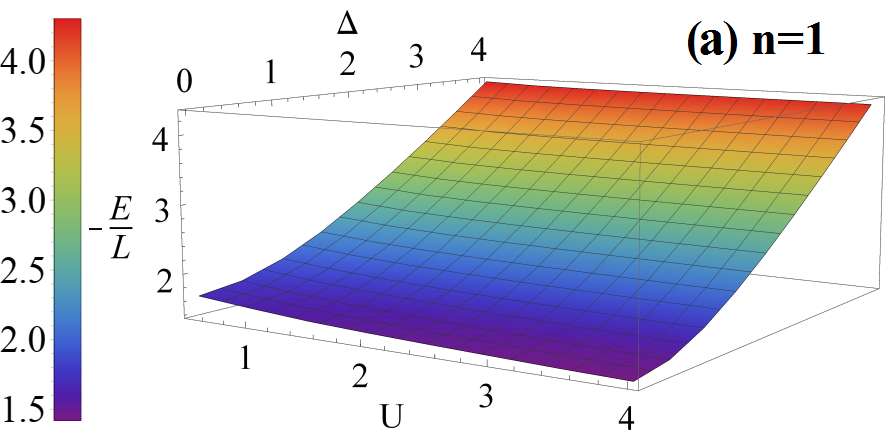} 
\includegraphics[scale=0.300]{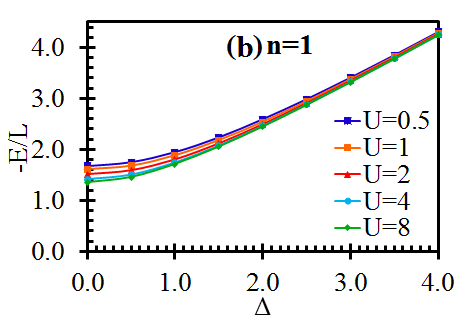}
\caption{Ground state energy of the quarter-filled IHM. Left panel shows the
dependence of the ground state energy on $U$ and $\Delta$ for $n=1$ electron per unit cell. Right panel
is the constant $U$ cuts of the same thing.}
\label{imbalance.fig}
\end{figure}

In Fig.~\ref{imbalance.fig} we have plotted the ground state energy as a function of $\Delta$ and $U$ for a fixed density
$n=1$ corresponding to quarter filling. As can be inferred from Eq.~\eqref{IHM.eqn}, the quantity $\partial E/\partial \Delta$ corresponds
to the average sublattice-imbalance, namely the difference in the average population of the two sublattices. 
As expected in the $\Delta=0$ limit, irrespective of the value of $U$, due to translational invariance the
average population of two sublattices must be the same. Therefore the slope of $E(\Delta)$ curve has to be zero. 
This can be clearly seen in both panels, and in particular in panel (b). For quite large values of $\Delta$, all curves
with Hubbard $U$ values upto $8.0$ converge to the same slope $\approx 1$. This simply means that when $\Delta$ is very large, 
one sublattice (with negative $\Delta$) is fully occupied, while the other sublattice is completely empty. 
Another interesting feature that can be seen in both (a) and (b) panels is that, for a constant finite $\Delta$ the
grounds state energy reduces by increasing the Hubbard $U$. This is better seen for smaller $\Delta$. This can be
explained as follows: In small $U$ limit and at quarter (or even less) filling the electrons dominantly occupy negative $\Delta$ sites
and positive $\Delta$ sites are empty. Due to kinetic term, electrons may come to positive $\Delta$ sites. Strong $U$ can take 
advantage of this and lower the ground state energy by superexchange mechanism. In this way the positive $\Delta$ sites will act like
effective ligand sites for the superexchange mechanism. As such, superexchange is suppressed by large $\Delta$.

\section{SUMMARY AND OUTLOOK}
In this work, in an attempt to find an exact solution, we generalized the Lieb-Wu solution 
of the standard Hubbard model to include the effect of non-zero ionic potential $\Delta$.
Every unit cell in IHM will have an additional sublattice label, $\eta=a,b$. This acts like an additional
orbital degree of freedom for every unit cell, albeit with different energies $\pm \Delta$. 
We constructed the two-particle scattering matrix in the space of $\eta$ and spin. For this purpose
we introduced extensions of the step function to properly define various "corners" in the configuration
space. The scattering matrix of the IHM satisfies the Yang-Baxter equation which indicates that our 
Bethe ansatz wave function, actually describes {\it some integrable model} (related to IHM). Even without knowing 
this model, our wave function is a highly non-trivial wave function the properties of which is worth
investigation.

Using our Bethe Ansatz wave function, we constructed the ground state energy as a function of 
various parameters $n,\Delta,U$. In the $\Delta=0$ limit, our results agree with the standard
Hubbard model. This wave function can be used to study various correlations in the ground state,
as well as the excitation spectrum with particular attention to the spin and charge excitations
and the effect of $\Delta$ in these excitations.

Since in every unit cell of the IHM there are two atomic orbitals with 
on-site energies $\pm\Delta$ (see Fig.~\ref{figure1}), by Choy-Haldane issue there will be
configurations in the Hilbert space which are not reachable by Bethe wave functions. 
In the limit of large $U$, and at low fillings, this issue becomes less important.
As such, our generalization of the Lieb-Wu wave function represents an effective 
solution of the IHM in the large $U$ and low-filling limit.  
Since the IHM has already been realized in a new solid state system based on twisted bilayer GeSe~\cite{Dante2019}, 
as well as in cold atom setting~\cite{Esslinger2015Ultracold} where $U$ is arbitrarily tunable, our effective exact
solution can be a useful tool to extract information about the properties of these systems.

In the absence of an analytic solution for IHM, our generalization of the Lieb-Wu wave function
is a promising playground to learn about the competition between the Hubbard $U$ and the ionic potential $\Delta$
in two directions: (i) An outstanding problem would be to search for solutions outside the family of Bethe wave functions 
(i.e. some sort of generalizations of our $\Delta$-dependent Bethe ansatz wave function) that are able to overcome the Choy-Haldane issue.
(ii) Although it would be nice to know the Hamiltonian to which our wave function corresponds, even without knowing 
what is the Hamiltonian, the {\it wave-function itself} contains a non-trivial deal of information about the
competition between two important insulating phases, namely the Mott and band insulators.  


\section*{ACKNOWLEDGMENTS}
This work was supported by grant number G960214 of the research deputy, Sharif University of Technology.
SAJ was supported by Iran Science Elites Federation (ISEF). 
We are grateful to \hlt{Natan} Andrei for many insightful comments. 
AH thanks R. Ghadimi for help in numerical computations. 
SAJ thanks Perimeter Institute of theoretical physics for hosting a visit during which this work 
was completed. 

\section*{APPENDICES}

\section*{Appendix A: Scattering matrix}

For two electrons, the wavefunction for the sublattice components is,
\begingroup\makeatletter\def\f@size{9.0}\check@mathfonts 
\begin{subequations} 
\bearr
F_{\sigma_1, \sigma_2}(x_1,b ; x_2,b)=\left(A_{\sigma_1, \sigma_2}\mathbb{H}(x_1- x_2)+B_{\sigma_1, \sigma_2}\mathbb{H}(x_2- x_1)\right)\nn\\
\times e^{i(k_1x_1+k_2x_2)}\chi_{b}(\theta_1)\chi_{b}(\theta_2)\nn\\
-\left(A_{\sigma_2, \sigma_1}\mathbb{H}(x_2- x_1)+B_{\sigma_2, \sigma_1}\mathbb{H}(x_1- x_2)\right)\nn\\
\times e^{i(k_1x_2+k_2x_1)}\chi_{b}(\theta_2)\chi_{b}(\theta_1),
\label{N2WFbb}
\eearr
\bearr
F_{\sigma_1, \sigma_2}(x_1,a ; x_2,a)=\left(A_{\sigma_1, \sigma_2}\mathbb{H}(x_1- x_2)+B_{\sigma_1, \sigma_2}\mathbb{H}(x_2- x_1)\right)\nn\\
\times e^{i(k_1x_1+k_2x_2)}\chi_{a}(\theta_1)\chi_{a}(\theta_2)\nn\\
-\left(A_{\sigma_2, \sigma_1}\mathbb{H}(x_2- x_1)+B_{\sigma_2, \sigma_1}\mathbb{H}(x_1- x_2)\right)\nn\\
\times e^{i(k_1x_2+k_2x_1)}\chi_{a}(\theta_2)\chi_{a}(\theta_1),
\label{N2WFaa}
\eearr
\bearr
F_{\sigma_1, \sigma_2}(x_1,b ; x_2,a)=\left(A_{\sigma_1, \sigma_2}[\mathbb{H}(x_1- x_2)+\frac{\delta_{x_1,x_2}}{2}]+B_{\sigma_1, \sigma_2}[\mathbb{H}(x_2- x_1)-\frac{\delta_{x_1,x_2}}{2}]\right)\nn\\
\times e^{i(k_1x_1+k_2x_2)}\chi_{b}(\theta_1)\chi_{a}(\theta_2)\nn\\
-\left(A_{\sigma_2, \sigma_1}[\mathbb{H}(x_2- x_1)-\frac{\delta_{x_1,x_2}}{2}]+B_{\sigma_2, \sigma_1}[\mathbb{H}(x_1- x_2)+\frac{\delta_{x_1,x_2}}{2}]\right)\nn\\
\times e^{i(k_1x_2+k_2x_1)}\chi_{b}(\theta_2)\chi_{a}(\theta_1),
\label{N2WFba}
\eearr
\bearr
F_{\sigma_1, \sigma_2}(x_1,a ; x_2,b)=\left(A_{\sigma_1, \sigma_2}[\mathbb{H}(x_1- x_2)-\frac{\delta_{x_1,x_2}}{2}]+B_{\sigma_1, \sigma_2}[\mathbb{H}(x_2- x_1)+\frac{\delta_{x_1,x_2}}{2}]\right)\nn\\ 
\times e^{i(k_1x_1+k_2x_2)}\chi_{a}(\theta_1)\chi_{b}(\theta_2)\nn\\
-\left(A_{\sigma_2, \sigma_1}[\mathbb{H}(x_2- x_1)+\frac{\delta_{x_1,x_2}}{2}]+B_{\sigma_2, \sigma_1}[\mathbb{H}(x_1- x_2)-\frac{\delta_{x_1,x_2}}{2}]\right)\nn\\
\times e^{i(k_1x_2+k_2x_1)}\chi_{a}(\theta_2)\chi_{b}(\theta_1).
\label{N2WFab}
\eearr
\end{subequations}
\endgroup

The Schr\"odinger equation for this wavefunction is given by the following matrix relation,
\begingroup\makeatletter\def\f@size{9.5}\check@mathfonts 
\begin{align}
\begin{pmatrix}
2\Delta +U\delta_{x_1,x_2} & -t\Delta_{2+} & -t\Delta_{1+} & 0 \\
-t\Delta_{2-} & 0 & 0 & -t\Delta_{1+} \\
-t\Delta_{1-} & 0 & 0 & -t\Delta_{2+} \\
0 & -t\Delta_{1-} & -t\Delta_{2-} & -2\Delta +U\delta_{x_1,x_2}
\end{pmatrix}&&
\begin{pmatrix}
F_{\sigma_1, \sigma_2}(x_1,b ; x_2,b)\\
F_{\sigma_1, \sigma_2}(x_1,b ; x_2,a)\\
F_{\sigma_1, \sigma_2}(x_1,a ; x_2,b)\\
F_{\sigma_1, \sigma_2}(x_1,a ; x_2,a)
\end{pmatrix}=\nn\\= \Delta(\cosh(\theta_1)+\cosh(\theta_2))
&&\begin{pmatrix} 
F_{\sigma_1, \sigma_2}(x_1,b ; x_2,b)\\
F_{\sigma_1, \sigma_2}(x_1,b ; x_2,a)\\
F_{\sigma_1, \sigma_2}(x_1,a ; x_2,b)\\
F_{\sigma_1, \sigma_2}(x_1,a ; x_2,a)
\end{pmatrix}
\end{align}
\endgroup
where 
\begingroup\makeatletter\def\f@size{8.8}\check@mathfonts 
\bearr
&&\Delta_{1\pm}F_{\sigma_1, \sigma_2}(x_1,\eta_1 ; x_2,\eta_2)=
F_{\sigma_1, \sigma_2}(x_1,\eta_1 ; x_2,\eta_2)+F_{\sigma_1, \sigma_2}(x_1\pm 1,\eta_1 ; x_2,\eta_2),\nn\\
&&\Delta_{2\pm}F_{\sigma_1, \sigma_2}(x_1,\eta_1 ; x_2,\eta_2)=
F_{\sigma_1, \sigma_2}(x_1,\eta_1 ; x_2,\eta_2)+F_{\sigma_1, \sigma_2}(x_1,\eta_1 ; x_2\pm 1,\eta_2).
\eearr
\endgroup

Then in the case $x_1=x_2 \equiv x$, for component $F_{\sigma_1, \sigma_2}(x,b ; x,b)$ we have,
\begingroup\makeatletter\def\f@size{9.0}\check@mathfonts 
\bearr
&&(2\Delta +U)F_{\sigma_1, \sigma_2}(x,b ; x,b)-t\Delta_{2+}F_{\sigma_1, \sigma_2}(x,b ; x,a)
-t\Delta_{1+}F_{\sigma_1, \sigma_2}(x,a ; x,b)\nn\\
&&=\Delta(\cosh(\theta_1)+\cosh(\theta_2))F_{\sigma_1, \sigma_2}(x,b ; x,b).
\eearr
\endgroup
By doing the calculations and using $\textbf{S}=\frac{1}{2}\left(\textbf{I}+\textbf{P}\right)+\frac{1}{2}\left(\textbf{I}-\textbf{P}\right)\textbf{\textit{s}}^{bb}$, the $\textbf{\textit{s}}^{bb}$ component of spin scalar becomes,
\begingroup\makeatletter\def\f@size{10.5}\check@mathfonts 
\be
\textbf{\textit{s}}^{bb}=\frac{\left(\frac{\chi_{a}(\theta_2)}{\chi_{b}(\theta_2)}\left(1-e^{ik_2}\right)-\frac{\chi_{a}(\theta_1)}{\chi_{b}(\theta_1)}\left(1-e^{ik_1}\right)\right)-\frac{U}{t}}{\left(\frac{\chi_{a}(\theta_2)}{\chi_{b}(\theta_2)}
\left(1-e^{ik_2}\right)-\frac{\chi_{a}(\theta_1)}{\chi_{b}(\theta_1)}\left(1-e^{ik_1}\right)\right)+\frac{U}{t}}.
\ee
\endgroup
Repeating these calculations for the $F_{\sigma_1, \sigma_2}(x,a ; x,a)$ component, the $\textbf{\textit{s}}^{aa}$ component of spin scalar obtain as,
\begingroup\makeatletter\def\f@size{10.5}\check@mathfonts 
\be
\textbf{\textit{s}}^{aa}=\frac{\left(\frac{\chi_{b}(\theta_1)}{\chi_{a}(\theta_1)}(1-e^{-ik_1})-\frac{\chi_{b}(\theta_2)}{\chi_{a}(\theta_2)}(1-e^{-ik_2})\right)-\frac{U}{t}}{\left(\frac{\chi_{b}(\theta_1)}{\chi_{a}(\theta_1)}(1-e^{-ik_1})-\frac{\chi_{b}(\theta_2)}{\chi_{a}(\theta_2)}(1-e^{-ik_2})\right)+\frac{U}{t}}.
\ee
\endgroup

\section*{Appendix B: Yang-Baxter Equation}
The Yang-Baxter equation or the star-triangle relation is a consistency equation which if it is established for a scattering matrix, the system (Hamiltonian) governing the dispersion will be integrable. In the case of our problem, this equation is written in the form,
\begingroup\makeatletter\def\f@size{10.0}\check@mathfonts 
\be \label{Yang-Baxter}
S_{1,2}(k_1,k_2)S_{1,3}(k_1,k_3)S_{2,3}(k_2,k_3)=S_{2,3}(k_2,k_3)S_{1,3}(k_1,k_3)S_{1,2}(k_1,k_2),
\ee
\endgroup
where the scattering matrix is,
\begingroup\makeatletter\def\f@size{10.5}\check@mathfonts 
\begin{equation}\label{Skt-Ap.B}
\begin{split}
\textbf{S}_{i,j}=\frac{\left(\alpha(k_i)-\alpha(k_j)\right)\textbf{I}_{i,j}+i\frac{U}{t}\textbf{P}_{i,j}}{\left(\alpha(k_i)-\alpha(k_j)\right)+i\frac{U}{t}}.
\end{split}
\end{equation} 
\endgroup
Assuming the two-particle scattering matrix at the brief form of $\textbf{S}_{i,j}\equiv \Upsilon_{ij}\textbf{I} +\Omega_{ij}\textbf{P}$, and by performing matrix multiplication the left-hand side(LHS) of Eq.~(\ref{Yang-Baxter}) is,
\begingroup\makeatletter\def\f@size{10.5}\check@mathfonts 
\begin{align}
LHS&=&(\Upsilon_{12}\Upsilon_{13}\Upsilon_{23}+\Upsilon_{12}\Omega_{13}\Omega_{23}+\Omega_{12}\Omega_{13}\Upsilon_{23}+\Omega_{12}\Upsilon_{13}\Omega_{23})\textbf{I}\nn\\
&&+(\Upsilon_{12}\Upsilon_{13}\Omega_{23}+\Omega_{12}\Upsilon_{13}\Upsilon_{23}+\Upsilon_{12}\Omega_{13}\Upsilon_{23}+\Omega_{12}\Omega_{13}\Omega_{23})\textbf{P}
\end{align}
\endgroup
Performing these calculations in a similar manner for the right-hand side(RHS) of Eq.~(\ref{Yang-Baxter}) indicates that RHS is exactly equal to LHS.


\bibliography{ref-son}

\end{document}